\documentclass{article}

\usepackage{microtype}
\usepackage{graphicx}
\usepackage{subfigure}
\usepackage{placeins}
\usepackage{booktabs} 

\usepackage{hyperref}

\usepackage[accepted]{icml2020}

\icmltitlerunning{Smart Car Features using Embedded Systems and IoT}

\begin{document}

\twocolumn[
\icmltitle{Smart Car Features using Embedded Systems and IoT}



\icmlsetsymbol{equal}{*}

\begin{icmlauthorlist}
\icmlauthor{Abhishek Das}{DJSCE}
\icmlauthor{Aditya Desai}{DJSCE}
\icmlauthor{Vivek Dhuri}{DJSCE}
\icmlauthor{Suyash Ail}{DJSCE}
\icmlauthor{Ameya Kadam}{DJSCE_Prof}
\end{icmlauthorlist}

\icmlaffiliation{DJSCE}{UG. Student, Department of Electronics and Telecommunication, DJSCE, Mumbai, India}
\icmlaffiliation{DJSCE}{UG. Student, Department of Electronics and Telecommunication, DJSCE, Mumbai, India}
\icmlaffiliation{DJSCE}{UG. Student, Department of Electronics and Telecommunication, DJSCE, Mumbai, India}
\icmlaffiliation{DJSCE}{UG. Student, Department of Electronics and Telecommunication, DJSCE, Mumbai, India}
\icmlaffiliation{DJSCE_Prof}{Assistant Professor, Department of Electronics and Telecommunication, DJSCE, Mumbai, India}



\vskip 0.3in
]



\printAffiliationsAndNotice{}  


\section{Abstract}
\label{submission}
There has been a tremendous rise in technological advances in the field of automobiles and autonomous vehicles. With the increase in the number of driven vehicles, the safety concerns with the same have also risen. The cases of accidents and life-threatening injuries have skyrocketed. It has become a necessity to provide adequate safety measures in automobiles. This project aims to develop a prototype for a smart vehicle system that provides real-time location of the vehicle on detection of a crash and alert the police station and relatives of the user, it has a panic button feature for a passenger’s safety. We also demonstrate a mechanism for cabin monitoring and an interactive interface between a user and a car, where the user can inquire about the temperature, humidity, and other variables inside the car remotely by sending a text message to the GSM module which is present in the car. The GSM module connects to the Arduino, which fetches the readings from sensors attached to it and sends it back to the user through a text message. We show the integration of MQ3 Alcohol sensor with Arduino for drunk driving prevention.

\section{Introduction}
\label{submission}

\begin{figure}
\vskip 0.2in
\begin{center}
\centerline{\includegraphics[width=\columnwidth]{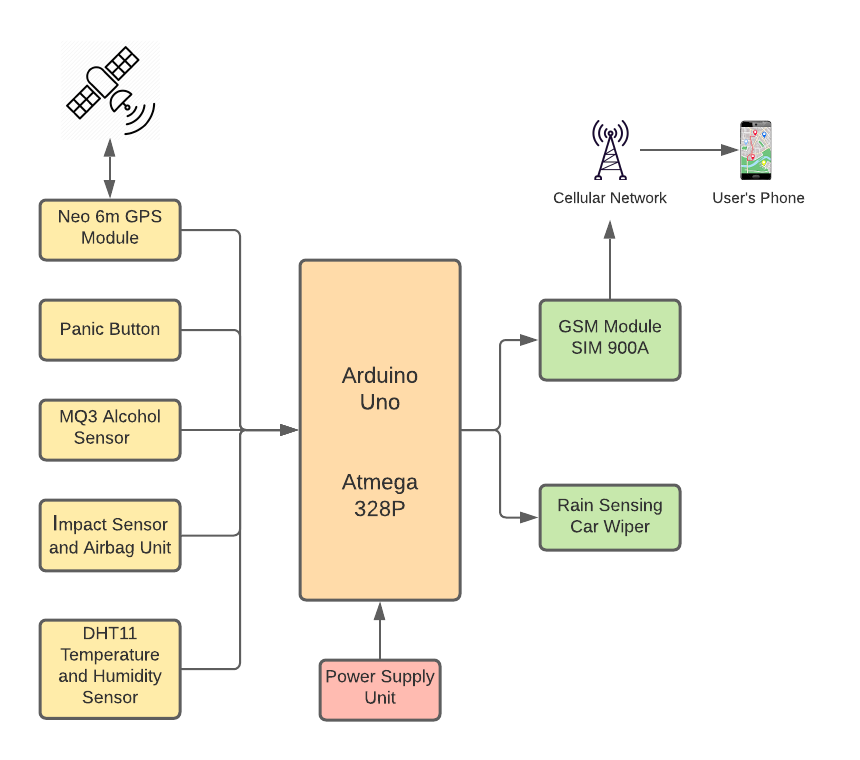}}
\caption{Block Diagram of Proposed Model}
\label{BlockDiagram}
\end{center}
\vskip -0.4in
\end{figure}

Driver safety has been an important feature in automobiles that have been made compulsion in various countries. An increasing number of amateur rash drivers, careless driving, and delayed access to first aid to victims has been a major cause of deaths. Cases of harassment, robbery in cabs are rising with more people using modern-day cab services. Driver fatigue monitoring, accident prevention measures, GPS-based location and nearest hospital alert, smart braking systems, smart airbags, etc. are some of the features currently implemented in a few of the high-end luxury-level vehicles. There hasn't been a cost-efficient model developed for the low-end budget cars. It is important to provide accessible safety measures in the vehicle to minimize the risk of loss of life. This project aims to develop a cost-efficient smart vehicle system that can help aid the cause. Figure \ref{BlockDiagram} shows the block diagram for our prototype. 

The primary objective of this project is to show how various sensors can be integrated to the Arduino or any microcontroller system, how to communicate with such a system remotely using technologies like GSM and GPS technology, send commands to inquire about the sensor readings and perform desired actions by using the actuators connected to the system. We have developed a low-cost prototype to demonstrate our ideas and create a baseline implementation for research purposes in this relatively new domain of the Internet of Things. With the recent developments in the capacity to process enormous amount of data from sensors, as well as communication technologies such as 5G, we believe our ideas can be scaled and deployed in real-time. 
The scripts have been made publicly available to the research community for further development
 \href{https://github.com/Abhishek0697/IoT_SmartCar}{here}.

In what follows, we discuss the related prior work for such a problem in the next section (3), followed by defining the Experimental Setup (4) and discussing our novel approaches in section (5), followed with its results and discussion in section (6). Finally, we end the discussion with conclusion and future directions in the last section (7).

\section{Related Work}

According to a Statistical Report \cite{indianexpress_2016} published by the Department of Roads and Highways Transport on Vehicle Mishaps in the country in 2016, the country has recorded 4,60,852 accidents in the year resulting in 1,45,685 deaths. Approximately 423 people died in 1,227 vehicle accidents every day. The data also states that at least 16 deaths occurring in vehicle mishaps out of 55 accidents in every hour in a particular period were primarily because victims were unable to receive suitable treatment within time. Thus, if an alert system is made and an alarm is raised, it might become possible to save many lives.

There has been prior work in the area of using GSM and GPS systems along with microcontrollers \cite{shinde2015real} developed a similar tracking system using an Embedded Linux board namely Raspberry Pi and a GSM SIM900A module.  The objective of their tracker was to raise an alert whenever the vehicle deviated from the predefined route which was set in the Raspberry Pi by the user. It also had features for sending notifications when the vehicle exceeded a set speed limit. \cite{saaid2014vehicle} implemented a vehicle location finder using a similar combination of GSM and GPS systems particularly for the task of vehicle thefts. 

\begin{figure}[t]
\vskip 0.2in
\begin{center}
\centerline{\includegraphics[width=6cm, height = 6 cm]{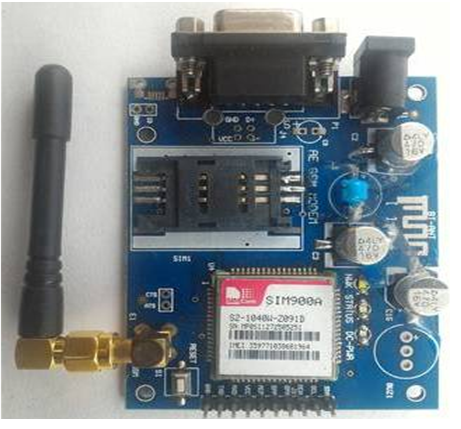}}
\caption{GSM SIM 900A module}
\label{GSMSIM900A}
\end{center}
\vskip -0.4in
\end{figure}

The use of panic buttons in vehicles is a idea which hasn't been deployed in real-life applications yet. According to the newspaper article \cite{hindustantimes_2016}, the Parliament of India will make it compulsory from 1st of April 2018 for all public transport vehicles which include buses and cabs to have a location tracker device and one or more panic buttons to alert the authorities in case of an emergency. Although, the government has not made the installation of cameras in these vehicles mandatory, primarily citing privacy concerns and due to the factor that it will generate tremendous amounts of data every second. The technology to process such huge data sets is currently unavailable. However, with the development in Internet of Things, in the future, this might be possible using Vehicle to Vehicle Communications. Nowadays, a large population of people chooses to travel by cabs and hence, keeping in mind the safety of the commuters, it is the need of the hour for developing such products. 

Another study by \cite{deutsche_welle} mentions that according to  National Crime Records Bureau (NCRB) report, drunk driving was a major factor in road accidents. 99 per cent of the fatal accidents that occur on the Highways are due to drunk driving and there is no check on this. Majority of these accidents involved trucks since the truck drivers drive irresponsibly when they are fully drunk. Until and unless the nation starts a new system of checking drunk driving on the highways, these fatalities cannot be reduced, as mentioned by a Joint Commissioner of Police. The current system of Drunk driver checking requires traffic police to make people blow into the breath-analyzers. However, it is not sufficient to check every instance of drunk driving cases due to the presence of an enormous number of vehicles on roads and especially outside cities and highways. Thus, an automatic monitoring system is needed to tackle this problem.

\begin{figure}
\vskip 0.2in
\begin{center}
\centerline{\includegraphics[width=7cm, height=6cm]{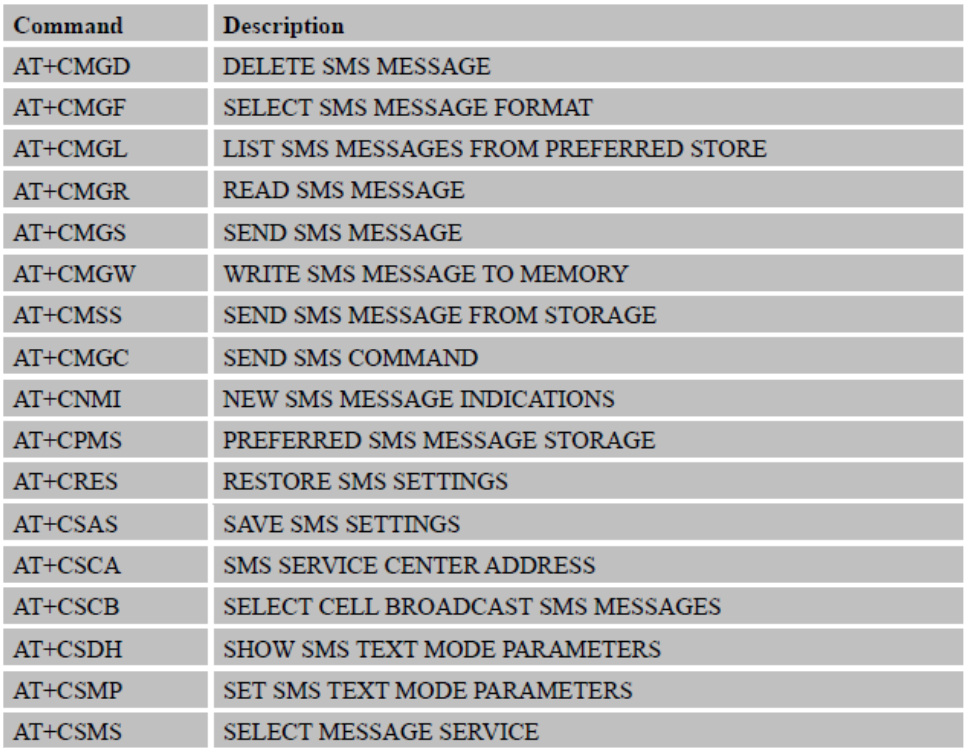}}
\caption{Basic AT commands used with GSM module}
\label{Basic}
\end{center}
\vskip -0.4in
\end{figure}

\section{Experimental Setup}

\subsection{GSM SMS Alert System}
SIM900A Modem is built with Dual Band GSM/ GPRS based SIM900A modem from SIMCOM. It works on frequencies 900/ 1800 MHz SIM900A can search these two bands automatically. The baud rate is configurable from 1200-115200 through AT command. This is a complete GSM module in an SMT type and designed with a very powerful single-chip processor integrating AMR926EJ-S core, allowing you to benefit from small dimensions and cost-effective solutions. Figure \ref{GSMSIM900A} shows a GSM SIM900A module.

\begin{figure}[t]
\vskip 0.2in
\begin{center}
\centerline{\includegraphics[width=4cm, height=4cm]{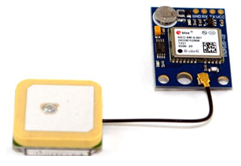}}
\caption{Neo 6m GPS module}
\label{GPS}
\end{center}
\vskip -0.4in
\end{figure}

\begin{figure}[t]
\vskip 0.2in
\centerline{\includegraphics[width=6cm, height = 4 cm]{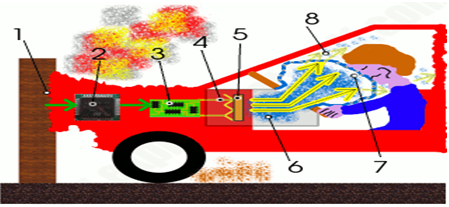}}
\caption{Triggering of Airbag circuitry and Accident Alert}
\label{Alert}
\vskip -0.in
\end{figure}

\subsection{GPS Tracking}
The NEO-6m module shown in Figure \ref{GPS} is a stand-alone GPS receiver featuring the high-performance u-blox 6 positioning engines. It is a flexible and cost-effective receiver that offers numerous connectivity options in a mini 160 x 122 x 24 cm package. It has a compact architecture and power and memory options which makes NEO-6m modules optimal for space constraint, low-cost devices. It has an acquisition engine, and 2 million effective correlators, and can make enormous parallel frequency searches, thus it can find a satellite within a small time. This 50-channel u-blox 6 positioning engine gives a Time to First Fix of around 1-2 seconds. It has an anti-jamming technology, Eeprom for storing settings which gives these receivers fantastic navigation performance even when placed in extremely difficult environments.

\subsection{Arduino Uno Development Board}
Arduino Uno is a development board based on a dual-inline-package ATmega328 AVR microcontroller \cite{mazidi20058051}. It has 20 digital input/ output pins, 6 of them can be used as Pulse Width Modulated \cite{holtz1992pulsewidth} outputs and 6 can be used as analog inputs. It has a 16 MHz crystal, a USB port, an ICSP header. Programs can be loaded onto it from the Arduino computer program software which is an open-source IDE. The Arduino has a vast support community, which makes it a very easy way to get started working with it.

\begin{figure}[t]
\vskip 0.2in
\begin{center}
\centerline{\includegraphics[width=6cm, height = 4 cm]{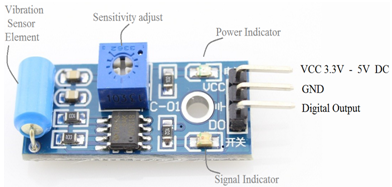}}
\caption{SW420 Impact Sensor  }
\label{SW420}
\end{center}
\vskip -0.4in
\end{figure}

\begin{figure}[t]
\vskip 0.2in
\begin{center}
\centerline{\includegraphics[width=4cm, height = 3.5 cm]{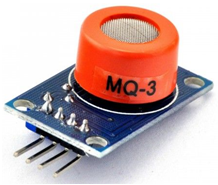}}
\caption{MQ3 Ethanol Sensor}
\label{MQ3}
\end{center}
\vskip -0.4in
\end{figure}

\section{Proposed Methodology}

\subsection{Accident Detection}

When a car hits something with a strong force, it starts to decelerate very rapidly. An Impact Sensor detects the change of velocity/amount of vibration. If the impact is great enough, the impact sensor triggers the airbag circuit and at the same time, it signals the Arduino to send an alert. Thus, when the impact is severe, the Arduino extracts the location by signalling the GPS module which connects with the GPS satellites and retrieves the location of the car. This location co-ordinates along with a google map link are sent to the designated mobile number in an SMS form through the GSM module. 

The SW420 sensor module gives outputs as ‘1’s or ‘0’s depending on vibration, tilt, and external force applied to it. In absence of vibration, this module gives logic ‘0’ as output and in presence of vibration, it gives logic ‘1’ as output. It has sensitivity control on the board. Figure \ref{SW420} shows the SW420 impact sensor which we have used in our prototype.

\subsection{Passenger Safety}
A panic button is placed such that whenever a passenger feels terror and discomfort due to certain reasons, an alert message is raised by sending an SMS on pressing the button. There can be multiple panic buttons placed at different spots in the vehicle and connected to the Arduino.

\subsection{Drunk Driver Prevention}
In this proposed system, an MQ-3 Ethanol Sensor as in Figure \ref{MQ3} is placed on the steering of the car or seat belt of the driver seat, such that it can monitor the percentage of alcohol in the breath of the driver. If it is found to be higher than set limits, then the Arduino signals the GSM to send an alert for the same to the driver’s predefined safety number (such as a home number). Measures can also include not to start the car engine unless the alcohol percentage reduces. When the user exhales, any ethanol present in their breath is oxidized to acetic acid. At the cathode, oxygen from the atmosphere is reduced. The overall reaction is the oxidation of ethyl alcohol. The charge flow produced by this reaction is measured and resistance is calculated, which results in the different levels of intoxication that the Arduino will determine.

\begin{figure}
\vskip 0.2in
\begin{center}
\centerline{\includegraphics[width=4cm, height=4cm]{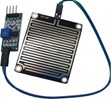}}
\caption{Rain Sensor Module}
\label{Rain Sensor Module}
\end{center}
\vskip -0.4in
\end{figure}

\subsection{Rain Sensing Automatic Wiper}
Car wipers in existing models are controlled manually by the driver. Some of the high-end cars have this feature, but due to cost factors, they have not yet made their way into normal vehicles. A cost-effective version of it is proposed in this project which includes a rain-drop sensor shown in Figure \ref{Rain Sensor Module} connected to the microcontroller, which is Arduino in this case. The rain sensor detects rain and sends the corresponding signal to the Arduino. This signal is then processed by the Arduino to take the desired action. The rain sensor consists of nickel tracks which when gets connected by water droplets in between the two tracks, the circuit gets connected and it detects rain. The raindrop sensor module is low cost and precise for raindrop detection. Its sensitivity can be changed by rotating the screw on the board. It has a digital output pin to indicate whether water is present or not and an analog output pin to give a measure of the intensity of water. The module has a power indicator led and separate control board.

A servo motor primarily contains a suitable motor, a gear reduction unit, a position measurement sensor, an control circuitry. Servo motor is a highly precise motor in terms of rotating angle. These are lightweight, low cost, compact motors that can be easily integrated into any circuits. The DC motor is connected to the gear unit which gives feedback to the position sensor. The potentiometer adjusts displacement according to the present position of the motor shaft. As the resistance changes, the differential voltage is generated. A PWM wave is given to the control wire which is transformed into voltage and is compared to the signal generated from the position sensor module. The control pin is connected to the Arduino’s PWM enabled pins.

\begin{figure}
\vskip 0.2in
\begin{center}
\centerline{\includegraphics[width=\columnwidth]{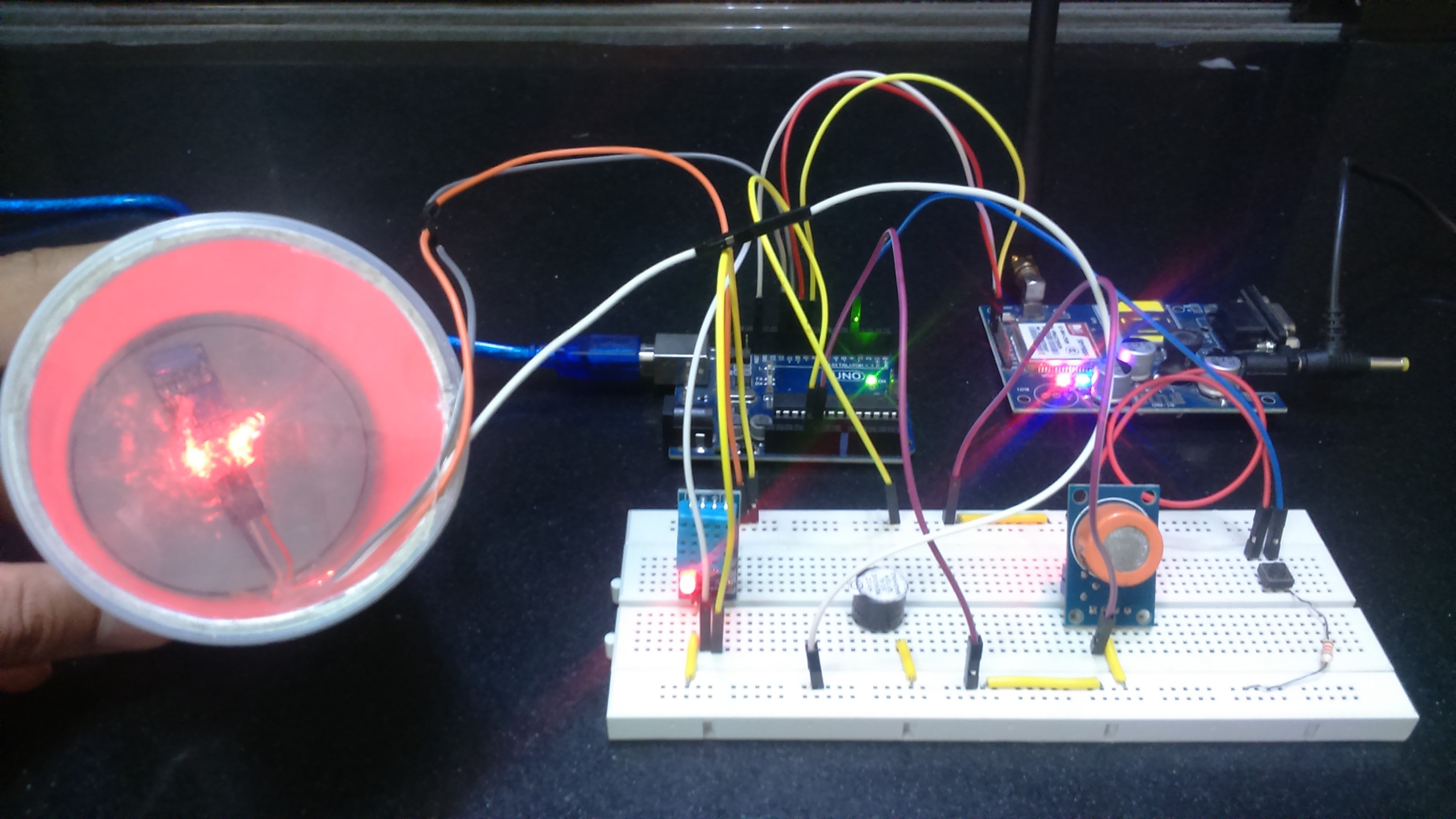}}
\caption{Demonstration of Prototype}
\label{model}
\end{center}
\vskip -0.2in
\end{figure}

\begin{figure}
\vskip 0.2in
\begin{center}
\centerline{\includegraphics[width=6cm, height = 9 cm]{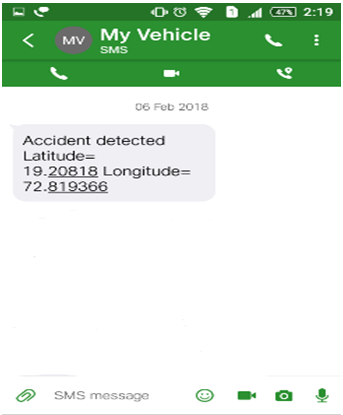}}
\caption{SMS sent for Accident detection}
\label{Alert}
\end{center}
\vskip -0.4in
\end{figure}

\section{Results and Discussion}

\begin{figure}
\vskip 0.2in
\begin{center}
\centerline{\includegraphics[width=6cm, height = 9 cm]{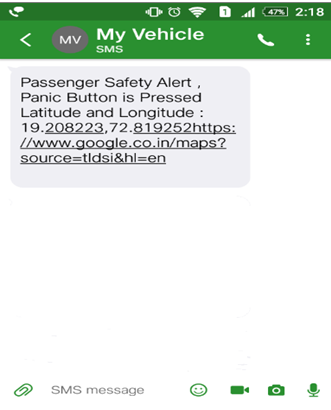}}
\caption{Panic Button Alert }
\label{Panic}
\end{center}
\vskip -0.4in
\end{figure}

Figure \ref{model} shows the prototype we have developed for demonstration. Figure \ref{Alert} shows the screenshot of an alert sent by our system on detection of an accident. The text message has information about the location coordinates of the car. The message could be sent to a police station or an designated relative by presetting the number in the system. Similarly, Figure \ref{Panic} shows the screenshot of the message delivered when the panic button in the vehicle is pressed. It has the location coordinated and the link to open it on the Google Map. In addition to these alerts, various other information can be sent in case of an emergency by modifying the code in the system. 

\section{Conclusion and Future Directions}
The sensors and equipment proposed in this prototype are low cost and efficient to a great extent, however in order to integrate this features with an existing vehicle's embedded system, a much more compact unit need to be built, which can be added on-chip. We believe that with the development of more high quality and accurate sensors, much more desirable and reliable outputs can be obtained.

The main purpose of our demonstrations was to put forward our research ideas with the hope of being further developed by the community and finally being scaled and deployed by the autonomous vehicle industry.

\FloatBarrier

\bibliography{example_paper}
\bibliographystyle{icml2020}

\end{document}